\documentclass[iop, apjl, tighten, bibyear]{emulateapj}
\pdfoutput=1
\usepackage{color}
\usepackage{amsmath}
\usepackage{amssymb}
\usepackage{enumitem}
\setlist{nolistsep}
\usepackage{graphicx}	% Including figure files
\usepackage{bm}		% Bold maths symbols, including upright Greek
\usepackage[plainpages=false, colorlinks=true, anchorcolor=blue, linkcolor=blue, citecolor=blue, bookmarks=false]{hyperref}
\usepackage[T1]{fontenc}
\usepackage{ae,aecompl}

\slugcomment{Accepted by the Astrophsysical Journal Letters, on April 27th 2016}

\shorttitle{ALMA resolves the torus of NGC\,1068}
\shortauthors{Garc\'{\i}a-Burillo et al.}

\begin{document}

%% LaTeX will automatically break titles if they run longer than
%% one line. However, you may use \\ to force a line break if
%% you desire.

\title{ALMA resolves the torus of NGC~1068: continuum and molecular line emission}
%
%% Use \author, \affil, and the \and command to format
%% author and affiliation information.
%% Note that \email has replaced the old \authoremail command
%% from AASTeX v4.0. You can use \email to mark an email address
%% anywhere in the paper, not just in the front matter.
%% As in the title, use \\ to force line breaks.
%
%%% Notice that each of these authors has alternate affiliations, which
%%% are identified by the \altaffilmark after each name.  Specify alternate
%%% affiliation information with \altaffiltext, with one command per each
%%% affiliation.
%
      \author{S.~Garc\'{\i}a-Burillo\altaffilmark{1}\altaffilmark{$\star$}, F.~Combes\altaffilmark{2}, C.~Ramos Almeida\altaffilmark{3,4}, A.~Usero\altaffilmark{1}, 
         M.~Krips\altaffilmark{5},  A.~Alonso-Herrero\altaffilmark{6}, S.~Aalto\altaffilmark{7},  V.~Casasola \altaffilmark{8},  L.~K.~Hunt
         \altaffilmark{8},  S.~Mart\'{\i}n\altaffilmark{9, 10}, S.~Viti\altaffilmark{11}, L.~Colina\altaffilmark{6,12},  F.~Costagliola \altaffilmark{7,13}, A.~Eckart \altaffilmark{14}, 
         A.~Fuente \altaffilmark{1}, C.~Henkel\altaffilmark{15,16},  I.~M\'arquez \altaffilmark{17}, R.~Neri	
         \altaffilmark{5}, E.~Schinnerer \altaffilmark{18}, L.~J.~Tacconi\altaffilmark{19}, \& P.~P.~van der Werf \altaffilmark{20}}                         
         \altaffiltext{$\star$}{\email[email: ]{s.gburillo@oan.es}}
         \altaffiltext{1}{Observatorio de Madrid, OAN-IGN, Alfonso XII, 3, 28014-Madrid, Spain} 
	 \altaffiltext{2}{LERMA, Obs. de Paris, PSL Research Univ., Coll\`ege de France, CNRS, Sorbonne Univ., UPMC, Paris, France}	
	 \altaffiltext{3}{IAC, V\'{\i}a L\'actea, s/n, E-38205 La Laguna, Tenerife, Spain}
	 \altaffiltext{4} {Departamento de Astrof\'{\i}sica, Universidad de La Laguna, E-38205, La Laguna, Tenerife, Spain}
	 \altaffiltext{5}{IRAM, 300 rue de la Piscine, Domaine Universitaire de Grenoble, 38406 
	 St.Martin d'H\`eres, France}	 
	   \altaffiltext{6}{CAB (CSIC-INTA), Ctra de Torrej\'on a Ajalvir, km 4, 28850 Torrej\'on de Ardoz, Madrid, Spain}	 	
	 \altaffiltext{7}{Department of Earth and Space Sciences, Chalmers University of Technology, Onsala Observatory, 439 92-Onsala, 
	 Sweden}
	 \altaffiltext{8}{INAF-Osservatorio Astrofisico di Arcetri, Largo Enrico Fermi 5, 50125-Firenze, Italy}
	  \altaffiltext{9}{Joint ALMA Observatory, Alonso de C\'ordova, 3107, Vitacura, Santiago 763-0355, Chile}
	   \altaffiltext{10}{ESO, Alonso de C\'ordova, 3107, Vitacura, Santiago 763-0355, Chile}	 
         \altaffiltext{11}{Department of Physics and Astronomy, UCL, Gower Place, London WC1E 6BT, UK}
          \altaffiltext{12}{ASTRO-UAM, Unidad Asociada CSIC, Madrid, Spain}         	 
	 \altaffiltext{13}{INAF - Istituto di Radioastronomia, via Gobetti 101, 40129, Bologna, Italy}
	 \altaffiltext{14}{I. Physikalisches Institut, Universit\"at zu K\"oln, Z\"ulpicher Str. 77, 50937, K\"oln, Germany}
	 \altaffiltext{15}{Max-Planck-Institut f\"ur Radioastronomie, Auf dem H\"ugel 69, 53121, Bonn, Germany}
	 \altaffiltext{16} {Astronomy Department, King Abdulazizi University, P.~O. Box 80203, Jeddah 21589, Saudi Arabia}	 	
	 \altaffiltext{17}{IAA (CSIC), Apdo 3004, 18080-Granada, Spain}
          \altaffiltext{18}{Max-Planck-Institut f\"ur Astronomie, K\"onigstuhl, 17, 69117-Heidelberg, Germany}	
	  \altaffiltext{19}{Max-Planck-Institut f\"ur extraterrestrische Physik, Postfach 1312, 85741-Garching, Germany}
	  \altaffiltext{20}{Leiden Observatory, Leiden University, PO Box 9513, 2300 RA Leiden, Netherlands}
%% Mark off your abstract in the ``abstract'' environment. In the manuscript
%% style, abstract will output a Received/Accepted line after the
%% title and affiliation information. No date will appear since the author
%% does not have this information. The dates will be filled in by the
%% editorial office after submission.

\begin{abstract}

We have used the Atacama Large Millimeter Array  (ALMA)  to map the emission of the CO(6--5) molecular line and the 432~$\mu$m continuum emission from the 300~pc-sized circumnuclear disk (CND) of the nearby  Seyfert~2 galaxy \object{NGC 1068} with a spatial resolution of $\sim4$~pc. These observations spatially resolve the CND and, for the first time, image the dust emission, the molecular gas distribution, and the kinematics  from a 7--10~pc-diameter disk that represents the submillimeter counterpart of the  putative  torus of \object{NGC\,1068}. 
We fitted the nuclear spectral energy distribution of the torus using ALMA and near and mid-infrared (NIR/MIR) data with {\sc CLUMPY} torus models. The  mass and radius of the best-fit solution for the torus are  both consistent with the values derived from the ALMA data alone: $M_{\rm gas}^{\rm torus}=(1\pm0.3)\times10^5~M_\sun$ and $R_{\rm torus}=3.5\pm0.5$~pc.  
 The dynamics of the molecular gas in the torus show strong non-circular motions and enhanced turbulence superposed on a surprisingly {\em slow} rotation pattern of the disk.  By contrast with the nearly edge-on orientation of the H$_2$O megamaser disk, we have found evidence suggesting that the molecular torus is less inclined ($i=34\degr-66\degr$) at larger radii. The lopsided morphology and complex kinematics of the torus could be the signature of the Papaloizou-Pringle instability, long predicted to likely drive the dynamical evolution of  active galactic nuclei (AGN) tori. 

\end{abstract}

%% Keywords should appear after the \end{abstract} command. The uncommented
%% example has been keyed in ApJ style. See the instructions to authors
%% for the journal to which you are submitting your paper to determine
%% what keyword punctuation is appropriate.


%% From the front matter, we move on to the body of the paper.
%% In the first two sections, notice the use of the natbib \citep
%% and \citet commands to identify citations.  The citations are
%% tied to the reference list via symbolic KEYs. The KEY corresponds
%% to the KEY in the \bibitem in the reference list below. We have
%% chosen the first three characters of the first author's name plus
%% the last two numeral of the year of publication as our KEY for
%% each reference.

%% Authors who wish to have the most important objects in their paper
%% linked in the electronic edition to a data center may do so by tagging
%% their objects with \objectname{} or \object{}.  Each macro takes the
%% object name as its required argument. The optional, square-bracket 
%% argument should be used in cases where the data center identification
%% differs from what is to be printed in the paper.  The text appearing 
%% in curly braces is what will appear in print in the published paper. 
%% If the object name is recognized by the data centers, it will be linked
%% in the electronic edition to the object data available at the data centers  
%%
%% Note that for sources with brackets in their names, e.g. [WEG2004] 14h-090,
%% the brackets must be escaped with backslashes when used in the first
%% square-bracket argument, for instance, \object[\[WEG2004\] 14h-090]{90}).
%%  Otherwise, LaTeX will issue an error. 

\section{Introduction}

\object{NGC~1068} is the prototypical Seyfert~2 galaxy and a prime example for active galactic nuclei (AGN) unifying schemes since the discovery of polarized optical continuum and broad-line emission in this source \citep{Mil83, Ant85}. Its central engine is thought to be hidden behind a screen of obscuring material located in a dusty molecular torus of a few parsecs size. Given the distance to the galaxy ($D\sim14$~Mpc; \citealt{Bla97}), which implies a scale $\sim$70~pc/$\arcsec$, spatially resolving the molecular/dusty torus has been challenging, as angular resolutions $<0\farcs1$ are required.

%%%%%%%%%%%%%%%%%%%%%%%% Figure  1
\begin{figure*}[ht!]
   \centering  
  \includegraphics[width=\textwidth]{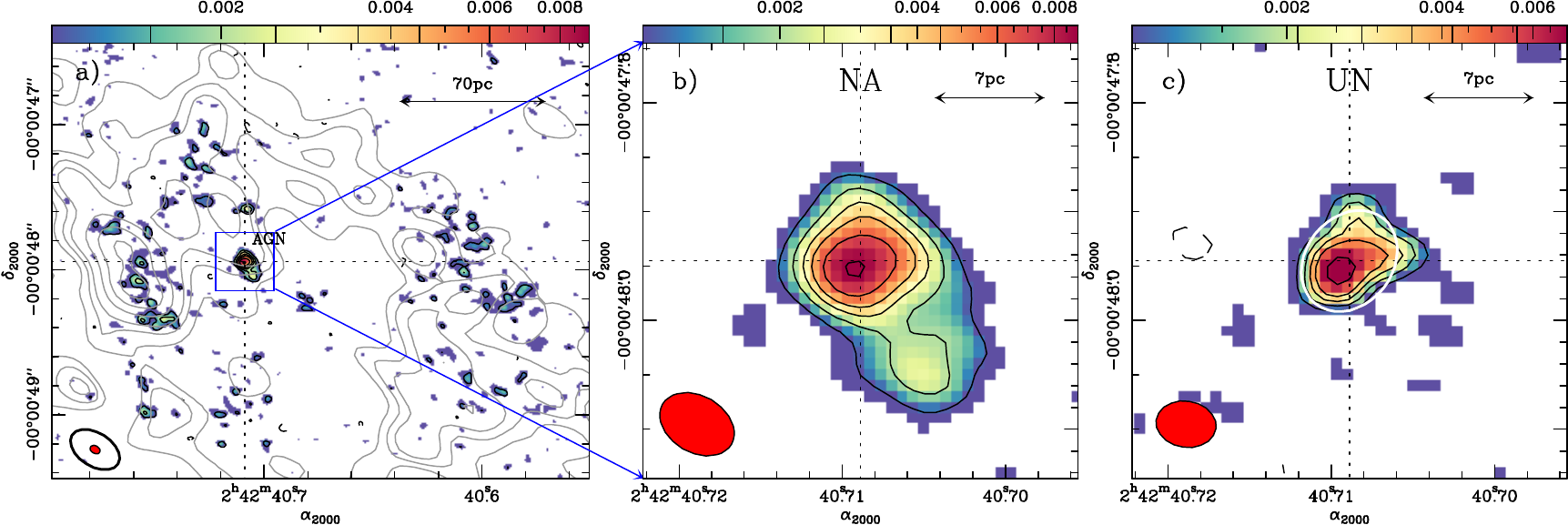}
\caption{(a)~The dust continuum emission at 694~GHz (432~$\mu m$) mapped by ALMA in the CND of NGC~1068. The natural (NA)-weighted map is 
shown in color scale (in Jy~beam$^{-1}$-units) and (black) contour levels  (3$\sigma$, 5$\sigma$, 7$\sigma$, 9$\sigma$, 12$\sigma$, 
and 16$\sigma$  where 1$\sigma=0.5$~mJy~beam$^{-1}$). The red-filled
ellipse at the bottom left corner represents the beam size at 694~GHz ($0\farcs07\times0\farcs05$ at  $PA=60^{\circ}$). Grey 
contours (10$\%$, 20$\%$, 30$\%$ to 90$\%$ in steps of 20$\%$ of the peak value: 49~mJy~beam$^{-1}$) identify the dust emission 
obtained in GB14 using ALMA with a lower resolution: $0\farcs4\times0\farcs2$ at  $PA=50^{\circ}$ (black ellipse).  The dashed lines highlight the location of the AGN at ($\alpha_{2000}, \delta_{2000}$) = ($02^{h}42^{m}40.709^{s}, -00^{\circ}00^{\prime}47.95\arcsec$).  (b)~A close-up of the dust continuum emission shown in left panel. %in the central $r\sim14$~pc around the 
%AGN. 
(c)~Same as middle panel but using a uniform (UN)-weighted set of data with a spatial resolution: $0\farcs06\times0\farcs04$ at  
$PA=82^{\circ}$. Contour levels  are 3$\sigma$, 4$\sigma$, 5$\sigma$, 6$\sigma$, and  8$\sigma$  where
1$\sigma=0.7$~mJy~beam$^{-1}$. The white ellipse identifies the disk solution found by the task {\tt UV\_FIT}. \label{Fig1}}
\end{figure*}

%%%%%%%%%%%%%%%%%%%%%%%%%

Single-dish observations in the near and mid-infrared (NIR and MIR) have shown extended emission around the central engine \citep{Boc00, Tom01, Rou04, Gal05, Gra06, Gra15}. Higher-spatial resolution interferometric observations in the NIR and MIR identified two components \citep{Jaf04,Wei04, Rab09, Bur13,Lop14}. First, a compact 0.5--1.4~pc-sized core at the AGN with a position angle $PA\simeq140^\circ$ that contains hot ($\simeq800$~K) dust co-spatial with the H$_2$O megamaser disk (\citealt{Gre96}, hereafter GR96; \citealt{Gre97}; \citealt{Gal01}, hereafter GA01; \citealt{Gal04}).

A second component of warm ($\simeq300$~K) dust extends over 3--10~pc along the north-south axis and may correspond to polar dust emission in the ionization cone. Due to their limited (u,v)-coverage, IR interferometric data have to be processed with reconstruction techniques to provide a best-fit model of the putative torus in the (u,v)-plane. This limitation can be to a large extent circumvented in the submillimeter using interferometers like ALMA, whose large number of antennas and baselines assure a much more complete coverage of the (u,v)-plane.

%%%%%%%%%%%%%%%%%%%%%%%%% Figure. 2 

\begin{figure}[tbh!]
\centering  
  \includegraphics[width=8.5cm]{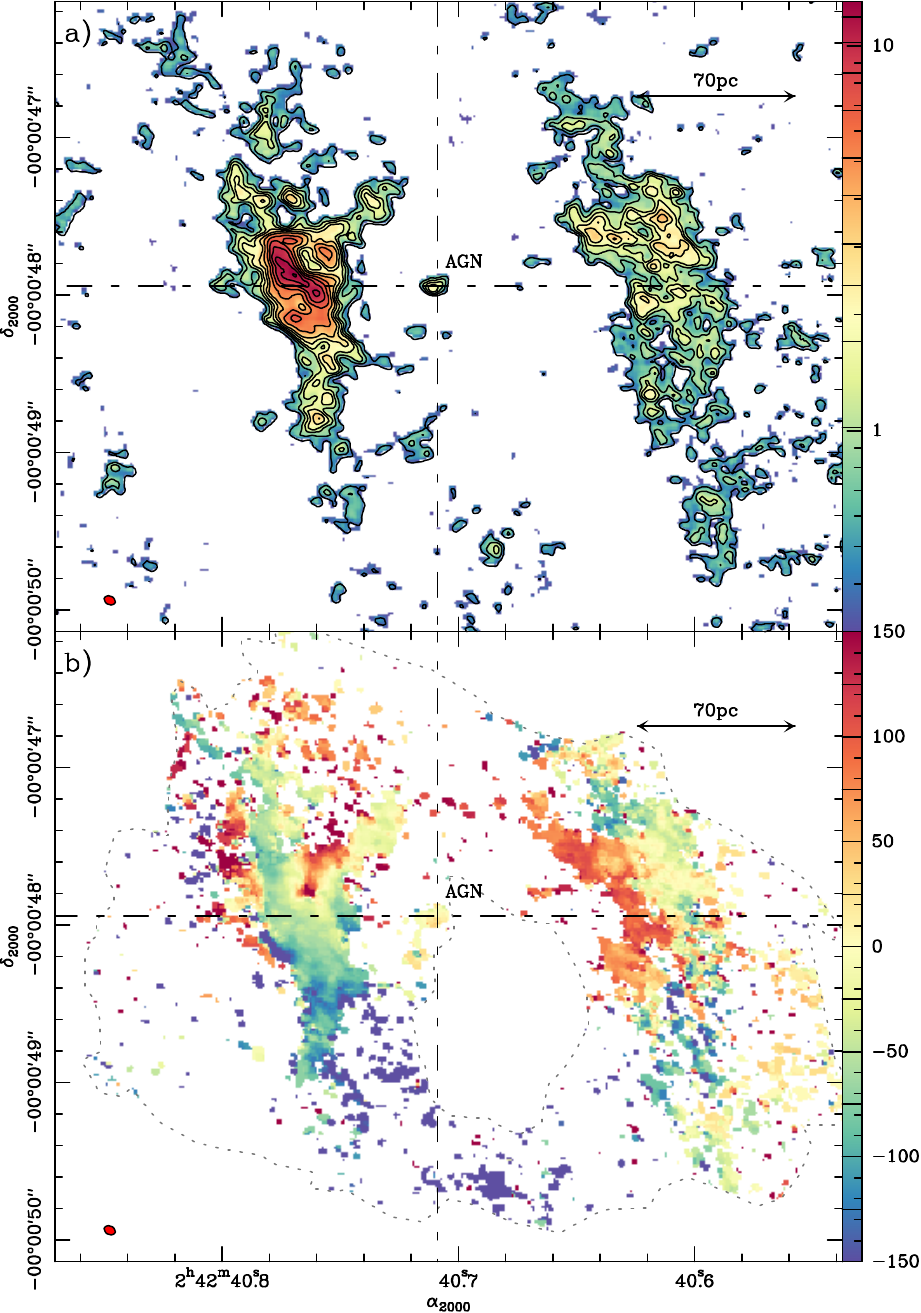}
\caption{(a)~The CO(6--5) intensity map of the CND of NGC~1068. The color scale and contours
span the range: 3$\sigma$, 5$\sigma$, 9$\sigma$, 12$\sigma$, 15$\sigma$ to  40$\sigma$ in steps of  5$\sigma$, 
where 1$\sigma=0.34$~Jy~km~s$^{-1}$beam$^{-1}$.  The filled ellipse at the bottom left corner represents the CO(6--5) beam size 
($0\farcs07\times0\farcs05$ at  $PA=60^{\circ}$). (b)~The CO(6--5) mean-velocity map (color scale). The dashed polygon identifies the region where significant ($>5\sigma$) emission was detected in GB14. Velocities refer to $v_{\rm o}$(HEL)$~=1136$~km~s$^{-1}$.  \label{Fig2}}
\end{figure}

%%%%%%%%%%%%%%%%%%%%%%%%% 

We used ALMA in Cycle 0 to image the dust continuum and  CO($J=6-5$) line emissions at 689~GHz in the circumnuclear disk (CND) of \object{NGC 1068} with a spatial resolution of 
$\sim20$~pc \citep[hereafter GB14]{Gar14}.  The CND appeared as a 300~pc$\times$200~pc-sized off-centered ellipsoidal ring with two prominent knots located 
east and west of the AGN. Although significant continuum and line emissions were also  detected at the position of the  AGN, the insufficient spatial resolution of these observations prevented us from isolating the torus. The new ALMA Cycle 2 observations of the CND of \object{NGC 1068} presented in this Letter have a spatial resolution of $\sim4$~pc. This factor of $\geq$20 smaller beam area compared to GB14, has allowed us to resolve the CND and image, for the first time, the dust emission and the distribution and kinematics of molecular gas from the torus of \object{NGC 1068} in the submillimeter wavelength range.

 %%%%%%%%%%%%%%%%%%%%%%%%% Fig 3

  \begin{figure}[tbh!]
\centering
\includegraphics[angle=0,width=7.5cm]{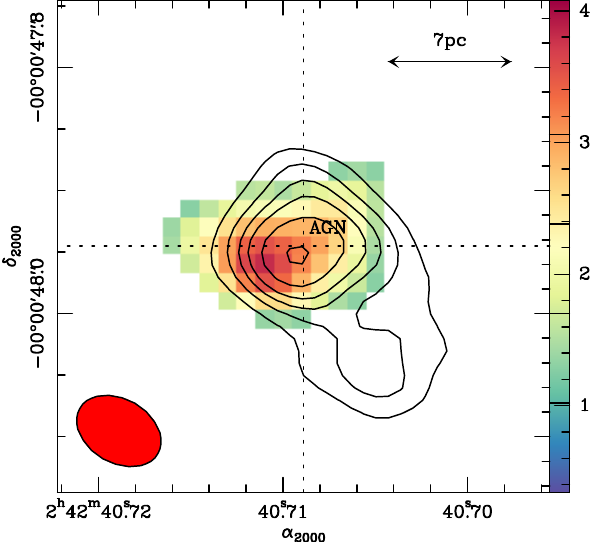}
\caption{
%(Upper panel)~The CO(6--5) emission stemming from the central $r\sim14$~pc around the 
%AGN. Most of the emission stems from a 7~pc--diameter disk. Color scale, contours, and symbols as in Fig.~\ref{Fig2}. (Lower panel
~Overlay 
of the continuum emission contours of Fig.~\ref{Fig1} on the CO(6--5) emission (color scale) from the AGN torus. Units are in Jy~km~s$^{-1}$beam$^{-1}$. %As in Fig.~\ref{Fig2}, the intersection of the dashed lines corresponds to the AGN position. \label{Fig3}} %Most of the continuum and line emissions stem from a 7-10~pc--diameter disk. 
\label{Fig3}}
\end{figure}

%%%%%%%%%%%%%%%%%%%%%%%%%
%
%\begin{figure}[tbh!]
%\centering
%\includegraphics[angle=0,width=7.5cm]{sed_ngc1068_midi_Y30.ps}
%\caption{SED fit}
%\end{figure}
%
%
%
%

 \section{ALMA observations and data reduction}

We observed the CO($J=6-5$) line and its underlying continuum emission (at 432~$\mu m$) in \object{NGC~1068} with ALMA 
during Cycle 2 using the Band~9 receiver (project-ID: $\#$2013.1.00055.S).  The projected baselines range from 41~m to 2251~m. The data were calibrated 
using the ALMA reduction package {\tt CASA} \citep{Mul07}. The calibrated uv-tables 
were exported to {\tt GILDAS\footnote{http://www.iram.fr/IRAMFR/GILDAS}} for mapping 
and CLEANing. One track was observed with 35 antennas during September 2015 using 
the extended configuration of  the array, which allowed us to reach a spatial resolution of 
$0\farcs07\times0\farcs05$ using natural (NA) weighting, and a field-of-view of $9\arcsec$ that covers the entire CND of the galaxy. Four spectral windows with a 
spectral bandwidth of 1.875~GHz were placed at sky frequencies of 687.970~GHz, 688.888~GHz,  
691.701~GHz, and 706.221~GHz. The second sub-band was centered at the redshifted frequency of the CO line.
%, which correspond, respectively, to the redshifted frequencies of the 
%H$^{13}$CN($J=8-7$),  CO($J=6-5$), SiO($J=16-15$), and HCN($J=8-7$) lines. While the CO line is clearly 
%detected, only upper limits are obtained for the other species. 
To optimize the deconvolution and cleaning we have used polygons to restrict the regions where cleaning components are identified. The polygons 
were defined for the CO line image for each channel based on the detection of significant ($\geq5\sigma$) 
emission identified in our Cycle 0 project  (GB14). The single polygon used in the cleaning of 
the continuum image was based on a similar detection criterion using the Cycle 0 data. 
%We assume a conservative accuracy of 
%$\sim30\%$ for the absolute flux scale. 
The sensitivity in the line data cube is 4~mJy~beam$^{-1}$  in channels of 12.8~km~s$^{-1}$ width. We obtained images of the continuum emission by averaging in each sub-band the channels free of possible line emission. The resulting maps were combined to produce an image  of the continuum emission  at 694~GHz. The corresponding point source sensitivity for the
continuum is 0.5~mJy~beam$^{-1}$.  Given the (u,v)-coverage, we expect to filter a significant 
amount of emission on scales $\geq2\arcsec$.
% for continuum emission. We expect that a comparatively lower fraction of the
%flux is filtered out in the CO line, due to the velocity structure of the emission. 

The phase tracking center of the central field is the same as in GB14 ($\alpha_{2000}, \delta_{2000}$) = ($02^{h}42^{m}40.771^{s}, -00^{\circ}00^{\prime}47.84\arcsec$) and the reference velocity is $v_{\rm o}$(HEL)$~=1136$~km~s$^{-1}$.

\section{Continuum and line emission maps}\label{maps}

---{\em The CND and the torus:} Figure~\ref{Fig1} shows the continuum emission of the CND imaged by the 
new ALMA observations.  While $\sim90\%$ of the total continuum flux is spread over a large number of faint emission clumps dispersed throughout the CND, the strongest emission peak corresponds, within $0\farcs02$ ($\sim1-2$~pc), to the position of the AGN (see Figs.~\ref{Fig1}ab).  The extended emission connecting the AGN with the CND in the map of GB14 is likely resolved out in the present data.  The AGN is identified as the source $S1$ in the radio-continuum maps of  \citet{Gal96} and  \citet{Gal04}\footnote{\citet{Ima16} have also recently isolated HCN(3--2) and HCO$^+$(3--2) emission from the AGN.}.

We have used the GILDAS task {\tt UV\_FIT} to find the best-fit to the visibilities inside a radius 
$r\leq0\farcs2$ (14~pc) from the AGN. The fit indicates that half of the total 
flux  in this region stems from a spatially resolved ellipsoidal  disk, characterized by a deconvolved diameter of 
$7\pm1$~pc oriented along  $PA^{\rm major}=142^{\circ}\pm23^{\circ}$.  Hereafter we refer to this fitted 
disk as the dust {\em torus}. The orientation of the dust torus is similar  to that of the H$_2$O maser  disk ($PA^{\rm maser}=140^{\circ}\pm5^{\circ}$; GR96, GA01). The  minor-to-major axis ratio of the dust torus  is $\simeq0.8\pm0.1$, an indication that the disk is spatially resolved along its minor axis. 
  Moreover, the residuals of the fit show  polar 
emission that extends south-west out to a distance of $\sim10$~pc relative to the torus major axis (Fig.~\ref{Fig1}b).  The enhanced spatial resolution of the uniformly-weighted data set better illustrates the goodness of the fit (Fig.~\ref{Fig1}c).

Figure~\ref{Fig2}a shows the CO(6--5) intensity map of the CND obtained by integrating 
the line emission in the velocity interval $v-v_{\rm o}=[-250, +250]$~km~s$^{-1}$.  The CO emission from the 
east and west knots of the CND is comparatively more prominent than in the continuum image  shown in  Fig.~\ref{Fig1}, which likely reflects the higher percentage of flux recovered for the line on scales $\geq2\arcsec$. 

Figure~\ref{Fig2}b shows the mean-velocity field of the gas  in the CND. The gas kinematics 
derived from the new observations are compatible with the picture drawn from GB14: gas motions in the CND  
show a superposition of outward radial flows on rotation. The molecular outflow
%, launched when the ionization cone of the NLR sweeps the disk in the CND, 
is responsible for tilting  the kinematic major axis by $\simeq70^{\circ}$ from $PA\simeq260^{\circ}$ in the outer galaxy disk to $PA\simeq330^{\circ}$ 
in the CND. 

Furthermore, ALMA detects the CO emission from a spatially-resolved lopsided disk located at the AGN, 
as illustrated in Fig.~\ref{Fig3}. The size (diameter$\simeq10\pm1$~pc, minor-to-major axis ratio $\simeq0.5\pm0.1$) and orientation ($PA^{\rm major}=112^{\circ}\pm20^{\circ}$) of the CO disk, hereafter referred to as the CO {\em torus}, are close to those derived above for the dust {\em torus}.  The orientation of the CO torus is virtually identical to that of the radio-continuum source $S1$ resolved in the VLBA map of \citet{Gal04}. Fig.~\ref{Fig3} shows no significant CO counterpart for the 
polar emission of the dust torus. %However, the observed gas kinematics indicate that the CO 
%torus is also spatially resolved along its minor axis (see Sect.~\ref{kin}). 

---{\em The mass of the torus:} From  the continuum flux of the dust torus at 432~$\mu m$, $\simeq13.8\pm1$~mJy, we estimate a dust mass of $M_{\rm dust}^{\rm torus}\sim1600~M_\sun$, 
assuming that a maximum of $\sim18\%$ of the flux at 694~GHz may come from other mechanisms different 
than thermal dust emission (GB14). To estimate $M_{\rm dust}^{\rm torus}$ we used a 
modified black-body function with a dust temperature $T_{\rm dust} \simeq T_{\rm gas}=150$~K,  an 
emissivity index $\beta=2$, and a value for the dust emissivity $\kappa_{{\rm 694~GHz}}$=0.34~m$^2$~kg$^{-1}$. The values of $T_{\rm dust}$ and $\kappa_{{\rm 694~GHz}}$ are based on the analysis of the line 
excitation and SED fitting done in GB14 and \citet{Vit14}, and on the adoption of a value for $\kappa_{{\rm 352~GHz}}=0.0865$~m$^2$~kg$^{-1}$ \citep{Kla01}. Assuming that the gas--to--dust ratio in the central 2~kpc 
of \object{NGC 1068} is $\simeq60$ (GB14), we derive a molecular gas mass for the torus  $M_{\rm gas}^{\rm 
torus}=(1\pm0.3)\times10^5~M_\sun$.  %The total gas mass derived for the central disk, including the torus and the 
%detected polar emission, is $\simeq(1.7\pm0.5)\times10^5~M_\sun$.

 %%%%%%%%%%%%%%%%%%%%%%%%%  Figure  4%%%%%%% 

\begin{figure}[tbh!]
\centering
\includegraphics[angle=0,width=8.5cm]{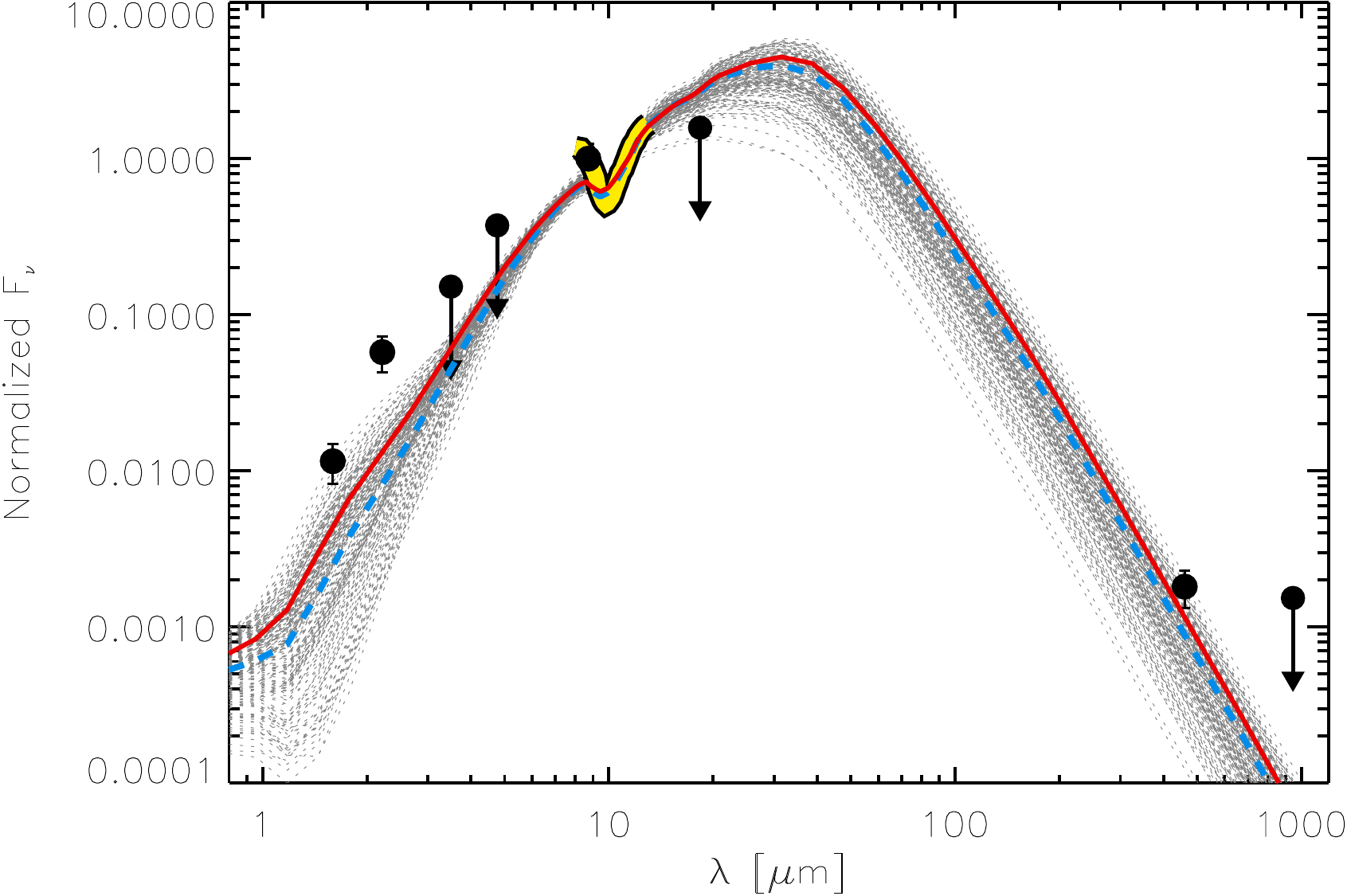}
\caption{High spatial resolution SED of NGC 1068 (thick yellow line: spectrum;
% sum of components 1 and 2 from 
%\citet{Lop14}
black dots: photometry) 
%\citep[this work]{Wei04, Mar00, Lop14, Tom01} 
normalized to the 8.7$\mu$m~point.
The solid red and dashed blue lines correspond to the maximum-a-posteriori (MAP) and median models respectively for the range
$60\degr<i<90\degr$. The upper limits correspond to the largest apertures ($\leq0\farcs3$). Grey curves are the CLUMPY models sampled from the posterior distributions compatible with the data at the 68$\%$ 
confidence level.\label{sedfit}}
\end{figure}

%%%%%%%%%%%%%%%%%%%%%%%%%

\section{Clumpy torus models}\label{clumpy}

We combined the ALMA Band 9 continuum thermal flux with the NIR/MIR continuum and MIR interferometry data obtained 
by several groups in apertures $\leq0\farcs05-0\farcs3$ to construct the nuclear SED of \object{NGC 1068} from $1.65\,\mu$m to $432\,\mu$m 
(\citealt{Wei04, Mar00, Lop14, Tom01}; this work). We also included as an upper limit the ALMA Cycle 0 Band 7 continuum flux 
from GB14. The 8--13 $\mu$m spectrum is the sum of components 1 and 2 in \citet{Lop14}. 
We note that this nuclear SED, shown in Fig.~\ref{sedfit}, corresponds to the smallest spatial scales probed to date for \object{NGC 1068}. 

We fitted the nuclear SED with the {\sc CLUMPY} torus models of 
\citet{Nen08a, Nen08b} adopting a Bayesian approach using the {\tt BayesClumpy} tool \citep{Ase09}. This scheme has been applied to model the SED of other Seyferts  \citep{Alo11, Ram11}. Using this approach we can specify a priori information about the six model
parameters. 
%We consider the priors to be truncated uniform distributions for the six model parameters 
%in the intervals reported in Table \ref{tab2}. 
We restricted the torus angular width $\sigma$ to be 40\degr$<\sigma<$60\degr~to match
the opening angle of the ionization cones (80\degr; \citealt{Das06}). The torus radial extent, defined by the outer--to--inner radius ratio $Y$, is in the range $5<Y<30$ based on the torus size measured  by ALMA at 432~$\mu m$.
%in the wavelength range probed here 
 In this first fit, we restricted  the inclination angle of the torus to $60\degr<i<90\degr$ based on the H$_2$O maser detection for this galaxy (GR96)
%The results of the fitting process are the posterior 
%distributions for the six parameters that describe the models 
%(defined in Table \ref{tab2}) 
%and the vertical shift required to match the fluxes to the observed SED.
We translate the results into the corresponding SED shown in Fig.~\ref{sedfit}. The NIR fluxes are under-predicted, an indication that we need to include some contribution from starlight, or a distinct component to reproduce the NIR bump (e.g.; \citealt{Kis12,Lir13}).

%The red solid line
%corresponds to the best-fitting model, described by the combination of parameters that maximize the %posterior (maximum-a-posteriori:~MAP). The blue dashed line represents the model computed with the %median value of each posterior parameter. 

%%%%%%%%%%%

%The {\sc CLUMPY} models reproduce the observed SED 
%with the exception of the 2.12 \micron~flux,  which might contain some stellar emission. 

%%%%%%%%%%%

%An estimation of the AGN bolometric luminosity can be obtained from the vertical shift applied 
%to the models to fit the data, which we allow to vary freely.  
%Using this shift, we obtain L$_{bol}^{AGN}$ = 1.6$\pm^{0.2}_{0.3}\times10^{44}~erg~s^{-1}$, a value that  
%can be compared with the bolometric luminosity estimated from the 2-10 keV luminosity. 

To compare our torus model with direct measurements from the ALMA observations presented here, 
we estimated from the fitted parameters the  gas mass $M_{\rm gas}^{\rm torus}$, 
%, which in turn is a function of 
%$\sigma$, $N_0$, $\tau_V$, $R_{\rm sub}$ and $Y$ 
%(see Section~6.1 in \citealt{Nen08b}),
the outer radius $R_{\rm torus}$, and the inclination $i$ of the torus. 
Using the median values of the fitted parameters, %reported in Table \ref{tab2}
we derive $M_{\rm gas}^{\rm torus} = 0.9\pm^{0.2}_{0.4} \times 10^{5}\,M_\odot$, $R_{\rm torus} = 4\pm$1 pc, and $i=66\degr\pm^{9}_{4}$.
These  values are consistent with those derived in Sect~\ref{maps} from the Band 9 observations alone. If we nevertheless allow the viewing angle of the torus to 
have smaller values in the range $30\degr<i<90\degr$ (see Sect.~\ref{kin}), the best-fit median values are  $M_{\rm gas}^{\rm torus} = 0.6\pm^{0.4}_{0.2} \times 10^{5}\,M_\odot$, $R_{\rm torus} = 2\pm$1 pc, and $i=34\degr\pm^{5}_{2}$.

%%%%%%%%%%%%%%%%%%%%%%%%%   Figure 5  %%%%%%%%
 
 \begin{figure*}[tbh!]
\centering
\includegraphics[angle=0,width=\textwidth]{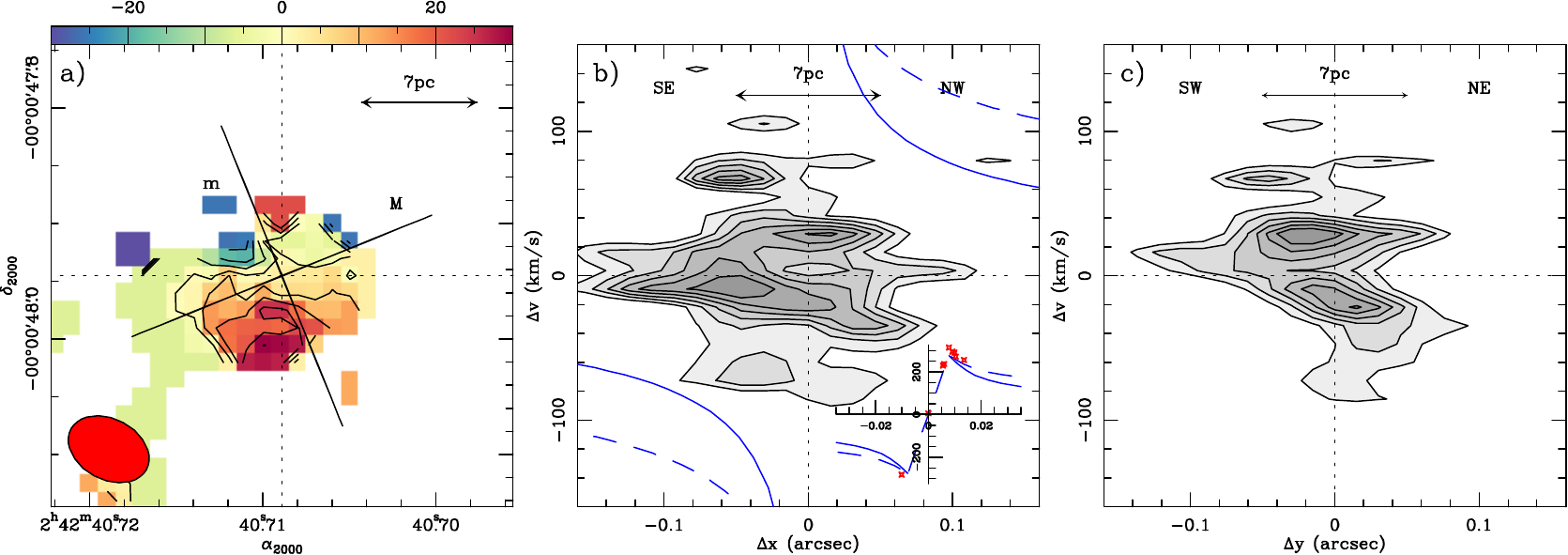}
\caption{(a)~The CO(6--5) mean-velocity map of the torus.  The contours span the range (--30~km~s$^{-1}$, 30km~s$^{-1}$) in steps of 10~km~s$^{-1}$. %The dashed blue line identifies the $PA$ of the dusty disk determined by ALMA ($PA^{\rm major}=145^{\circ}\pm23^{\circ}$). 
The black lines labeled as $M$ and $m$ show, respectively, the orientations of the major and minor axes of the CO torus  ($PA^M=112^{\circ}\pm20^{\circ}$, $PA^m=22^{\circ}\pm5^{\circ}$). (b)~The CO(6--5) position-velocity diagram along the $M$ axis.  Contours go from 
2$\sigma$ to  7$\sigma$ in steps of  1$\sigma$, where 1$\sigma=4$~mJy~$^{-1}$beam$^{-1}$. The inset shows the velocities relative to $v_{\rm sys}$ as a function of radius (in arcseconds) 
% $v_{\rm o}$(HEL)$~=1150$~km~s$^{-1}$ 
as derived for the H$_2$O megamaser spots (red markers) detected along  $PA^{\rm maser}=140^{\circ}\pm5^{\circ}$ (GR96, GA01). The dashed (solid) blue curve shows the best-fit sub-Keplerian (Keplerian) rotation curve  
$v_{\rm rot}\propto r^{-\alpha}$  of GR96 with $\alpha=0.31$ (0.50). (c)~Same as middle panel but for the $m$ axis. In all panels, velocities refer to $v_{\rm o}$(HEL)$~=1136$~km~s$^{-1}$ and ($\Delta x$,$\Delta y$)--offsets are relative to the AGN position. \label{Fig5}}
\end{figure*}

 %%%%%%%%%%%%%%%%%%%%%%%%%

%The  parameters in the {\sc CLUMPY} models are the torus size $Y$  defined as the ratio
%between the outer radius $R_{\rm o}$ and the inner radius\footnote{The inner radius is set by the assumed dust 
% sublimation temperature of 1500~K and the AGN bolometric
% luminosity.} $R_{\rm
%  sub}$, the torus angular size $\sigma$, the viewing
%angle $i$, the number of clouds along the equatorial direction $N_0$,
%the optical depth of the clouds $\tau_V$, and the index $q$ of the
%radial distribution of the clouds $\propto r^{-q}$. Using a
%Bayesian approach to fit the data with the {\tt BayesClumpy} tool (Asensio Ramos \& Ramos Almeida~\cite{Ase09}) the torus model
%parameters of \object{NGC~1068} 

\section{Kinematics of the torus}\label{kin}

Figure~\ref{Fig5} shows that ALMA has spatially resolved the molecular gas kinematics in the torus. 
 Although the mean-velocities derived from CO, shown in Fig.~\ref{Fig5}a, span a moderate range ($v-v_{\rm o}=[-30, +30]$~km~s$^{-1}$), we identify spatially-resolved velocity gradients within the torus.
 Overall, the line emission appears redshifted to the south and blueshifted to the north. Therefore, the apparent {\em kinematic} major axis of the CO disk, at $PA^{\rm kin}\simeq180^{\circ}$, is tilted by a significant angle ($\simeq53^{\circ}\pm15^{\circ}$)
 %% average btw=38=180-142 degs (dust) and 68 degs =180-112 (co). Note Gratadour=118 degs%%%%
 relative to the {\em morphological} major axis of the CO/dust torus, oriented at
 a weighted average $PA^{\rm major}\simeq127^{\circ}\pm15^{\circ}$ (see Sect.~\ref{maps}). 
 % average btw=112 degs (co) and 142 degs (torus). Note Gratadour=118 degs%%%%
Taken at face value, the tilt between both axes indicates that strong non-circular motions are superposed to the {\em  slow} rotation pattern of the CO torus.
  
Figure~\ref{Fig5}b shows the position-velocity (p-v) diagram taken along the morphological major axis of the CO torus  
 ($PA^M=112^{\circ}\pm20^{\circ}$). The bulk of the CO emission in the torus spans a wide range of velocities: $v-v_{\rm o}\simeq[-80,80]$~km~s$^{-1}$. Figure ~\ref{Fig5}b compares 
 the observed kinematics with the velocity range expected from the H$_2$O sub-Keplerian rotation curve  
 of \citet{Gal96} projected along the CO major axis ($v_{\rm rot}\propto v^{-\alpha}$, with $\alpha=0.31$; see also \citealt{Lod03}). Even in the most favorable case of a purely Keplerian  extrapolation ($\alpha=0.5$), the CO velocities are still a factor $\geq2$ lower compared to the H$_2$O-based kinematics for an edge-on disk ($i=90\degr$). To reach a better agreement we would need to incline the CO torus down to $i=30\degr-40\degr$, i.e., close to the inclination of the galaxy disk (GB14). Although these small values 
 for $i$ can be discarded for the {\em mostly} edge-on maser disk (at $r\leq1$~pc), we cannot rule out that the CO torus is tilted towards a more face-on orientation (at $r\geq2-4$~pc). Furthermore, this scenario is still within the range of solutions found in Sect.~\ref{clumpy} for the clumpy torus. However, as argued below, the intricate kinematics of the CO torus cannot be attributed alone to the changing orientation of a purely rotating disk. 
 
 Figures~\ref{Fig5}a,b show that the CO emission  is redshifted on the southeast side of the major axis at $\Delta x\simeq-0\farcs06$. The emission is nevertheless blueshifted northwest at $\Delta x\simeq+0\farcs06$. This velocity gradient, measured on scales $\Delta x\simeq\pm4$~pc from the AGN, is inverted relative to that measured for the H$_2$O maser disk at $\Delta x\simeq\pm1$~pc. The apparent counter-rotation of the outer disk relative to the inner disk could be caused by an enhancement of the gas turbulence  in the torus. A rough estimate of the ratio of the deprojected mean velocity ($v_{\rm mean}$) to the velocity  dispersion ($\sigma$) of the gas along the major axis yields very low values: $v_{\rm mean}/\sigma<1$.   However, Fig.~\ref{Fig5}c shows that a comparable spatially-resolved velocity gradient is measured along the minor axis p-v diagram. The tilted disk scenario or the simple addition of turbulence to rotation cannot alone explain the observed pattern along the minor axis,  an indication that non-circular motions in the torus need to be invoked. 
 
\citet{Pap84} first predicted that non self-gravitating AGN tori should likely develop low-order non-axisymmetric instabilities on a dynamical time-scale. This type of instability, known as the Papaloizou-Pringle instability (PPI), has been studied in isolated self-gravitating tori \citep{Kiu11, Kor13}, and also in tori perturbed by the accretion of matter coming from the outer boundary of the system \citep{Don14}. These new simulations show the growth of a long-lasting $m=1$ mode related to the onset of the PPI. The $m=1$ mode generates a lopsided distribution, as well as enhanced turbulence and strong non-circular motions in the torus gas. Evidence of inflowing gas connecting the CND with the torus has been found in the  2.12~$\mu m$ H$_2$ lines by \citet{Mue09}. The remarkably lopsided distribution and complex kinematics of the $\sim7-10$~pc CO torus of \object{NGC~1068} seen by ALMA could be the footprint of the PPI in this Seyfert. Alternatively, the velocity gradient observed along the minor axis could be explained by gas being entrained in the outflow, as similarly detected by 
\citet{Cec02} in high-ionization lines close to the AGN. This radial shift is noticeably reversed farther out at the CND where the CO outflow follows the pattern of low-ionisation lines (GB14).

\section{Summary and conclusions}

We have mapped with ALMA the dust continuum at 432~$\mu m$ and the CO(6--5) molecular line emission in the CND of \object{NGC 1068} with a spatial resolution of $\sim4$~pc. 
These observations have allowed us to resolve the CND and image the dust emission and, also, the distribution and kinematics of molecular gas from a 7--10~pc-diameter disk. This is the first detection of a submillimeter counterpart of the  putative torus of \object{NGC 1068}. As expected for the contribution of a comparatively cooler component, the CO/dust torus extends on spatial scales that are twice larger compared to the brightest MIR sources detected by the VLTI at the position of the central engine of \object{NGC 1068} \citep{Bur13,Lop14}. 

The ALMA measurements have been combined 
with the NIR/MIR continuum and MIR interferometry data available 
%obtained in apertures $\leq0\farcs05-0\farcs3$ 
from $1.65\,\mu$m to $432\,\mu$m. We fitted the nuclear SED with the {\sc CLUMPY} torus models of \citet{Nen08a, Nen08b}. The  mass and the radius of the best-fit torus are consistent with the values derived from the ALMA data alone: $M_{\rm gas}^{\rm torus}=(1\pm0.3)\times10^5$~M$_\sun$ and $R_{\rm torus}=3.5\pm0.5$~pc.

We have spatially resolved the kinematics of molecular gas in the torus. The overall {\em slow} rotation pattern of the disk is perturbed by strong non-circular motions and enhanced turbulence. %The kinematical major axis of the CO torus is tilted relative to the morphological major axis. 
Furthermore,  while the maser condition of the H$_2$O line suggests that the inner molecular disk should be oriented mostly edge-on at $r\leq1$~pc, we have found evidence that the molecular torus could be tilted towards lower inclination angles ($i=34\degr-66\degr$) at  $r\geq2-4$~pc. The lopsided morphology and complex kinematics of the \object{NGC~1068} torus could be the signature of the PPI instability predicted by \citet{Pap84}. This instability was foreseen to rule the dynamical evolution of AGN tori.  %A detailed comparison with the predictions of the PPI scenario will be the subject of a forthcoming paper.

The results of this work are relevant to caution against the use of CO lines as a direct tracer of the supermassive black hole mass ($M_{\rm BH}$) in galaxies (e.g.; \citealt{Dav13}). In cases like \object{NGC 1068}  where even mid-J CO lines trace the outer boundary of a kinematically distorted tilted torus, careful modeling would be mandatory to reliably estimate $M_{\rm BH}$.

\acknowledgments
SGB, AU, LC, and IM  acknowledge support from Spanish grants AYA2012-32295 and AYA2013-42227-P. We thank the referee, Makoto Kishimoto, for his constructive review.
This paper makes use of the following ALMA data: ADS/JAO.ALMA$\#$2013.1.00055.S. ALMA is a partnership of ESO (representing its member states), NSF (USA) and NINS (Japan), together with NRC (Canada), NSC and ASIAA (Taiwan), and KASI (Republic of Korea), in cooperation with the Republic of Chile. The Joint ALMA Observatory is operated by ESO, AUI/NRAO, and NAOJ.

\end{document}